\documentclass[preprint,12pt,sort&compress]{elsarticle}
\usepackage{amsfonts}
\usepackage{amsmath}
\usepackage{amsthm}
\usepackage{amssymb}
\usepackage{graphicx}
\usepackage{dcolumn}
\usepackage{bm}
\usepackage{bbding}
\usepackage{mathrsfs}
\usepackage{graphicx}

\biboptions{numbers,sort&compress}

\begin{document}

\begin{frontmatter}
\title{Deterministic transfer of an unknown qutrit state assisted by the low-$Q$ microwave resonators}
\author{Tong Liu}
\author{Yang Zhang}
\author{Chang-Shui Yu\corref{cor1}}
\ead{quaninformation@sina.com}
\cortext[cor1]{Corresponding author}
\author{Wei-Ning Zhang \corref{cor2}}
\address{School of Physics and Optoelectronic Technology, Dalian University of
Technology, Dalian 116024, China}
\begin{abstract}
Qutrits (i.e., three-level quantum systems) can be used to achieve many quantum information and communication tasks due to their large Hilbert spaces.
In this work, we propose a scheme to transfer an unknown quantum state between two flux qutrits coupled to two superconducting
coplanar waveguide resonators. The quantum state transfer can be deterministically achieved without measurements. Because resonator photons
are virtually excited during the operation time, the decoherences caused by the resonator decay and
the unwanted inter-resonator crosstalk are greatly suppressed. Moreover, our approach can be adapted to other solid-state qutrits coupled to circuit resonators. Numerical simulations show that the high-fidelity transfer of
quantum state between the two qutrits is feasible with current circuit
QED technology.
\end{abstract}

\end{frontmatter}

\section{Introduction}

Circuit quantum electrodynamics (QED), consisting of
microwave resonators and superconducting (SC) qubits (i.e., two-level quantum systems), is considered as one of the most promising
candidates for quantum information processing (QIP) \cite{s2,s3,s4,s5} and quantum simulations \cite{Nori1,Nori2,Nori3,Nori4}. SC qubits have been employed
as a testing platform for quantum computation and QIP due to their significantly increased coherence times, controllability, scalability, and interfaceability
\cite{s6,s7,s8,s9,s10,s11,s12,Pop,s13}. SC qubits based on Josephson junctions are mesoscopic element circuits like ``artificial atoms", with multiple discrete energy levels whose spacings can be rapidly adjusted by varying
external control parameters (e.g., magnetic flux applied to the
superconducting loop of a superconducting phase, transmon,
Xmon, or flux qubit; see, e.g.,~\cite{s8,s14,s15,s16}). In circuit QED, the third level of SC artificial atom has already
been used to implement a number of fundamental tasks in QIP~\cite{s12,You1,You2,You3,s17,s18,s19,s20,s21,K. S. Kumar},
and the leakage errors of a SC qutrit have been efficiently reduced in experiments~\cite{Z. J. Chen}. Furthermore, the
preparation of a ground state~\cite{K. Geerlings,X. Y. Jin} or an arbitrary three level superposition state~\cite{s22} for a SC qutrit has also been demonstrated.

On the other hand, quantum state transfer is an essential primitive in QIP and quantum communication. Over the past years, a great many of theoretical proposals have been proposed for realizing SC qubit-to-qubit quantum state transfer based on cavity/circuit QED \cite{s23,yangprl,s24,s26,s27,s28,s29,s30,yangnjp,yangprb}. Moreover,
the quantum state transfer between the SC qubits has been experimentally demonstrated in circuit QED. For instance, Refs.~\cite{s31,s32,s33} experimentally demonstrated
quantum state transfer between two SC qubits through a resonator. Ref.~\cite{s34} realized the SC qubit-to-qubit quantum state transfer assisted with SC microwave resonators. In these proposals, the resonator acts as a quantum data bus
to mediate long-range and fast interaction between SC qubits.

The transmission of an unknown quantum state between two solid-state qutrits is still an open problem in QIP. Hitherto, the previous works are mainly focused on the quantum state transfer from one qubit to another qubit~\cite{s23,yangprl,s24,s26,s27,s28,s29,s30,yangnjp,yangprb,s31,s32,s33,s34}.
Few proposals are related to quantum state transfer between qutrits and there is no experimental
study is reported for the qutrit-to-qutrit quantum state transfer in circuit QED.
Recently, the probabilistic transfer of qutrit state
between particles via a spin chain has been presented \cite{14Bayat}, and a method has been proposed for deterministic transfer of a qutrit state assisted by the high-$Q$ microwave resonators \cite{s44}.
In this paper, we propose a scheme to achieve an unknown quantum state transfer between two flux qutrits coupled to two SC
coplanar waveguide resonators~(Fig.~1). The three levels of qutrit $j$ are denoted as $|g\rangle_j,$ $|e\rangle_j$ and $|f\rangle_j$ $(j=1,2)$ [Fig.~2]. The quantum state transfer from qutrit 1 to qutrit 2 is expressed as
\begin{eqnarray}
(\alpha|g\rangle_1+\beta|e\rangle_1+\gamma|f\rangle_1)~|g\rangle_2
\rightarrow|g\rangle_1~(\alpha|g\rangle_2+\beta|e\rangle_2+\gamma|f\rangle_2),
\end{eqnarray}
where $\alpha,$ $\beta$ and $\gamma$ are the normalized complex numbers; the subscripts 1 and 2 represent qutrits 1 and 2.

This proposal has the following features and advantages: (\romannumeral1)
The quantum state can be deterministically transferred without measurements;
(\romannumeral2) Because the resonator photons
are virtually excited for during the operation time, the decoherences caused by the resonator decay and
the unwanted inter-resonator crosstalk are greatly suppressed; (\romannumeral3) This approach is quite
general because it can be applied to accomplish the same task with other solid-state qutrits coupled to circuit resonators; (\romannumeral4) Through the numerical analysis, we discuss the possible experimental implementation of our proposal and show that the average fidelity could reach 99.2\%.

This paper is organized as follows. In Sec.~2, we show how to
transfer an unknown state between two SC flux qutrits. In Sec.~3, we discuss the possible experimental implementation of our
proposal and numerically calculate the operational fidelity for transferring
the state from flux qutrit 1 to qutrit 2. A concluding summary is given
in Sec. 4.

\section{Quantum state transfer between two qutrits}

\begin{figure}[tbp]
\begin{center}
\includegraphics[bb=63 457 515 804, width=8.5 cm, clip]{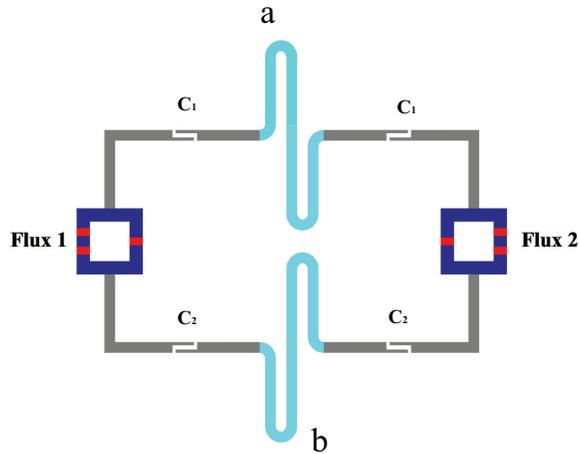} \vspace*{%
-0.08in}
\end{center}
\caption{Setup of two flux qutrits (blue squares) coupled to two resonators
via capacitances $C_1$ and $C_2$. The
flux qutrits can be other types of solid-state qutrit, such as a quantum dot or a superconducting phase qutrit.}
\label{fig:1}
\end{figure}

\begin{figure}[tbp]
\begin{center}
\includegraphics[bb=10 420 577 665, width=12.5 cm, clip]{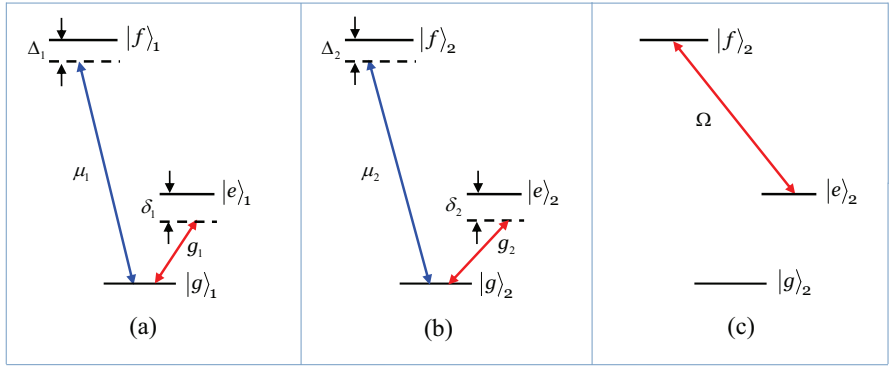}
\vspace*{-0.08in}
\end{center}
\caption{(a) Resonator $a$ $(b)$ is far-off resonant with $|g\rangle\leftrightarrow|e\rangle$ ($|g\rangle\leftrightarrow|f\rangle$) transition of qutrit 1 with
coupling strength $g_1$ ($\mu_1$) and detuning $\delta_1$ ($\Delta_1$).
(b) Resonator $a$ $(b)$ is far-off resonant with $|g\rangle\leftrightarrow|e\rangle$ ($|g\rangle\leftrightarrow|f\rangle$) transition of qutrit 2 with
coupling strength $g_2$ ($\mu_2$) and detuning $\delta_2$ ($\Delta_2$). (c) The microwave pulse is resonant with the $|e\rangle\leftrightarrow|f\rangle $ transition of qutrit 2, with a Rabi frequency $\Omega$.}
\label{fig:2}
\end{figure}

Consider a system which consists of two flux qutrits connected by two resonators [Fig.~1]. As shown in Fig.~2 (a,b), resonator $a$ ($b$) is off-resonantly coupled to the
$|g\rangle\leftrightarrow|e\rangle $ ($|g\rangle\leftrightarrow|f\rangle $) transition of qutrit $j$ with a coupling constant $g_{j}$ ($\mu_{j}$). Here, $j=1,2$. Suppose that qutrit 1 is initially in an unknown state $(\alpha|g\rangle_1+\beta|e\rangle_1+\gamma|f\rangle_1)$, qutrit 2 is in the state $|g\rangle_2$ and two
resonators are in the vacuum state $|0\rangle_a|0\rangle_b$.
In the interaction picture, the
Hamiltonian of the whole system can be written as (in units of $\hbar =1$)
\begin{eqnarray}
H_{I,1}=\sum_{j=1}^2 g_{j}(e^{i\delta_j t}a\sigma
_{eg,_j}^{+}+h.c.)
+\sum_{j=1}^2 \mu_{j}(e^{i\Delta_j t}b\sigma_{fg,_j}^{+}+h.c.),
\end{eqnarray}
where $\sigma
_{eg,_j}^{+}=|e\rangle_j\langle g|$, $\sigma
_{fg,_j}^{+}=|f\rangle_j\langle g|$, $\delta_j=\omega_{eg,_j}-\omega_{a}$, and $\Delta_j=\omega_{fg,_j}-\omega_{b}$ ($j=1,2$).  Here, $\omega_{eg,_j}$ ($\omega_{fg,_j}$) is the $|g\rangle\leftrightarrow|e\rangle $ ($|g\rangle\leftrightarrow|f\rangle $) transition frequency of qutrit $j$ and $\omega_{a}$ ($\omega_{b}$) is the frequency of resonator $a$ ($b$).

Consider the large detuning conditions $\delta _{j}\gg g_{j}$ and $\Delta_j\gg\mu_{j}$.
It is straightforward to show that the Hamiltonian (2) changes to (for details, see Ref. \cite{s35})
\begin{eqnarray}
H_{eff} &=&\sum_{j=1}^2\frac{g_{j}^2}{\delta _{j}}(aa^\dagger|e\rangle_j\langle e|-a^\dagger a|g\rangle_j\langle g|)
+\sum_{j=1}^2\frac{\mu_{j}^2}{\Delta _{j}}(b b^\dagger|f\rangle_j\langle f|-b^\dagger b|g\rangle_j\langle g|)\nonumber \\
&+&\sum_{j=1}^2\frac{g_j\mu_j}{2}(\frac{1}{\delta _{j}}+\frac{1}{\Delta_{j}})~[~a^\dagger b\sigma_{fe,_j}^{+}e^{-i(\delta_j-\Delta_j)t}+ h.c.~]\nonumber \\
&+&\lambda_1~[e^{i(\delta_1-\delta_2) t}\sigma_{eg,_1}^{+}\sigma_{eg,_2}^{-}+ h.c.]+\lambda_2~[e^{i (\Delta _{1}-\Delta _{2})t}\sigma_{fg,_1}^{+}\sigma_{fg,_2}^{-}+ h.c.],
\end{eqnarray}%
where $\lambda_1=\frac{g_1g_2}{2}(\frac{1}{\delta _{1}}+\frac{1}{\delta _{2}})$ and $\lambda_2=\frac{\mu_1\mu_2}{2}
(\frac{1}{\Delta _{1}}+\frac{1}{\Delta _{2}})$.
As mentioned previously, resonators $a$ and $b$ are initially in the vacuum state, and we set
\begin{eqnarray}
\delta _{1}=\delta _{2}=\delta, \Delta _{1}=\Delta _{2}=\Delta.
\end{eqnarray}%
Then the Hamiltonian (3) reduces to
\begin{eqnarray}
H_{eff}=H_{0}+H_{i},
\end{eqnarray}
with
\begin{eqnarray}
H_{0}=\sum_{j=1}^2\frac{g_{j}^2}{\delta _{j}}|e\rangle_j\langle e|
+\sum_{j=1}^2\frac{\mu_{j}^2}{\Delta _{j}}|f\rangle_j\langle f|,
\end{eqnarray}
\begin{eqnarray}
H_{i}=\lambda_1(\sigma_{eg,_1}^{+}\sigma_{eg,_2}^{-}+ h.c.)+\lambda_2(\sigma_{fg,_1}^{+}\sigma_{fg,_2}^{-}+ h.c.),
\end{eqnarray}
Note that the Hamiltonians (6) and (7) do not contain the operators of the resonators. Thus, each resonator remains in the vacuum state.

In a new interaction picture under the Hamiltonian $H_0$ and we set
\begin{eqnarray}
g_{1}^2/\delta _{1}=g_{2}^2/\delta _{2}, \mu_{1}^2/\Delta_{1}=\mu_{2}^2/\Delta _{2},
\end{eqnarray}
that is $g_1=g_2=g$ and $\mu_{1}= \mu_{2}=\mu$ according to Eq. (4), which can be achieved by a prior design of the sample with appropriate capacitances $C_1$ and $C_2$. Thus, one can obtain
\begin{eqnarray}
\widetilde{H}_{i}=e^{iH_0t}H_ie^{-iH_0t}={H_{i}}.
\end{eqnarray}
In addition, we set
\begin{eqnarray}
g^2/\delta=\mu^2/\Delta,
\end{eqnarray}
one can obtain $\lambda_1=\lambda_2=\lambda$. Therefore, the Hamiltonian (7) can be expressed as
\begin{eqnarray}
\widetilde{H}_{i}=\lambda~[(\sigma_{eg,_1}^{+}\sigma_{eg,_2}^{-}+ h.c.)+(\sigma_{fg,_1}^{+}\sigma_{fg,_2}^{-}+ h.c.)].
\end{eqnarray}
Based on Hamiltonian (11) and after returning to the original interaction picture by
performing a unitary transformation $e^{-iH_0t}$, the following state evolution can be obtained
\begin{eqnarray}
|g\rangle_1|g\rangle_2&\rightarrow&|g\rangle_1|g\rangle_2, \nonumber\\
|e\rangle_1|g\rangle_2&\rightarrow&e^{-i\lambda t}\cos(\lambda t)|e\rangle_1|g\rangle_2-ie^{-i\lambda t}\sin(\lambda t)
|g\rangle_1|e\rangle_2\nonumber \\
|f\rangle_1|g\rangle_2&\rightarrow&e^{-i\lambda t}\cos(\lambda t)|f\rangle_1|g\rangle_2-ie^{-i\lambda t}\sin(\lambda t)
|g\rangle_1|f\rangle_2.
\end{eqnarray}
When the evolution time $t$ is equal to $\pi/2\lambda$, one obtains the transformations
$|g\rangle_1|g\rangle_2\rightarrow|g\rangle_1|g\rangle_2$, $|e\rangle_1|g\rangle_2\rightarrow-|g\rangle_1|e\rangle_2$,
$|f\rangle_1|g\rangle_2\rightarrow-|g\rangle_1|f\rangle_2$ simultaneously. Consequently, we have the
following state transformation
\begin{eqnarray}
(\alpha|g\rangle_1+\beta|e\rangle_1+\gamma|f\rangle_1)~|g\rangle_2
\rightarrow|g\rangle_1~(\alpha|g\rangle_2-\beta|e\rangle_2-\gamma|f\rangle_2),
\end{eqnarray}

Adjust the level spacings of each qutrit such that it is decoupled from the two resonators. Then apply a microwave pulse to qutrit 2. The pulse is resonant with the $|e\rangle\leftrightarrow|f\rangle$ transition of qutrit 2~[Fig.~2(c)].
The Hamiltonian in the interaction picture is written as
\begin{eqnarray}
H_{I_,2}=\Omega(|e\rangle _{2}\langle
f|+h.c.),
\end{eqnarray}
where $\Omega$ is the Rabi frequency of the microwave pulse. One obtains the following rotations under the
Hamiltonian~(14),
\begin{eqnarray}
|e\rangle _{2} &\rightarrow &\cos (\Omega t)|e\rangle _{2}-i\sin (\Omega t)|f\rangle _{2},  \nonumber \\
|f\rangle _{2} &\rightarrow &\cos (\Omega t)|f\rangle _{2}-i\sin (\Omega t)|e\rangle _{2}.
\end{eqnarray}
We set $t_2=\pi/\Omega$ to obtain a $\pi$ rotation by $|e\rangle _{2}\rightarrow-|e\rangle _{2}$ and $|f\rangle _{2}\rightarrow-|f\rangle _{2}$. Hence, we can obtain $-|g\rangle_1|e\rangle_2\rightarrow|g\rangle_1|e\rangle_2$ and $-|g\rangle_1|f\rangle_2\rightarrow|g\rangle_1|f\rangle_2$. Hence, it follows from Eq.~(13)
\begin{eqnarray}
(\alpha|g\rangle_1+\beta|e\rangle_1+\gamma|f\rangle_1)|g\rangle_2
\rightarrow|g\rangle_1(\alpha|g\rangle_2+\beta|e\rangle_2+\gamma|f\rangle_2),
\end{eqnarray}
which shows that the original superposition state of qutrit 1 is perfectly transferred to qutrit 2 after the above operation.

\section{Possible experimental implementation}

When the inter-resonator crosstalk is taken
into account, the Hamiltonian~(2) becomes
$\widetilde{H}_{I,1}=H_{I,1}+\epsilon$, where $\epsilon$ describes the unwanted inter-resonator crosstalk,
given by $\epsilon=g_{ab}( e^{i\Delta_{ab} t}ab^{+}+h.c.)$, with the inter-resonator crosstalk coupling strength
$g_{ab}$ and the two-resonator frequency detuning $\Delta_{ab} =\omega_{b}-\omega_{a}$.
In addition, we also consider the inter-resonator crosstalk coupling during the qutrit-pulse resonant interaction. Therefore, the Hamiltonian~(14) is modified as
$\widetilde{H}_{I,2}=H_{I,2}+\epsilon$.

When the dissipation and the dephasing are included, the dynamics of the lossy system is determined by the
following master equation
\begin{eqnarray}
\frac{d\rho }{dt} &=&-i[ \widetilde{H}_{I},\rho ] +\kappa _{a}
\mathcal{L}[a]+\kappa _{b}\mathcal{L}[b]\nonumber \\
&+&\sum_{j=1,2}\left\{ \gamma _{eg_j}\mathcal{L}[ \sigma
_{eg,_j}^{-}] +\gamma _{fe_j}\mathcal{L}[ \sigma _{fe,_j}^{-}]
+\gamma _{fg_j}\mathcal{L}[ \sigma _{fg,_j}^{-}] \right\}  \nonumber\\
&+&\sum_{j=1,2}\left\{ \gamma _{j,\varphi f}\left( \sigma _{ff_j}\rho
\sigma _{ff_j}-\sigma _{ff_j}\rho /2-\rho \sigma _{ff_j}/2\right) \right\} \nonumber \\
&+&\sum_{j=1,2}\left\{ \gamma _{j,\varphi e}\left( \sigma _{ee_j}\rho
\sigma _{ee_j}-\sigma _{ee_j}\rho /2-\rho \sigma _{ee_j}/2\right) \right\},
\end{eqnarray}
where $\widetilde{H}_{I}$ is $\widetilde{H}_{I,1}$ or $\widetilde{H}_{I,2}$ above, $\sigma _{eg,_j}^{-}=\left\vert g\right\rangle _{j}\left\langle
e\right\vert$, $\sigma _{fe,_j}^{-}=\left\vert e\right\rangle _{j}\left\langle
f\right\vert$, $\sigma _{fg,_j}^{-}=\left\vert g\right\rangle _{j}\left\langle
f\right\vert , \sigma _{ee_j}=\left\vert e\right\rangle _{j}\left\langle
e\right\vert ,\sigma _{ff_j}=\left\vert f\right\rangle _{j}\left\langle
f\right\vert;$ and $\mathcal{L}\left[ \Lambda \right] =\Lambda \rho \Lambda
^{+}-\Lambda ^{+}\Lambda \rho /2-\rho \Lambda ^{+}\Lambda /2,$ with $\Lambda=a,b,\sigma _{eg,_j}^{-},\sigma _{fe,_j}^{-},\sigma _{fg,_j}^{-}.$ Here, $\kappa
_{a}$ ($\kappa_{b}$) is the photon decay rate of resonator $a$ ($b$). In addition, $
\gamma _{eg_j}$ is the energy relaxation rate of the level $\left\vert
e\right\rangle $ of qutrit $j$, $\gamma _{fe_j}$ ($\gamma _{fg_j}$) is the
energy relaxation rate of the level $\left\vert f\right\rangle $ of qutrit $
j $ for the decay path $\left\vert f\right\rangle \rightarrow \left\vert
e\right\rangle $ ($\left\vert g\right\rangle $), and $\gamma _{j,\varphi e}$
($\gamma _{j,\varphi f}$) is the dephasing rate of the level $\left\vert
e\right\rangle $ ($\left\vert f\right\rangle $) of qutrit $j$ ($j=1,2$).

The fidelity of the operation is given by
$\mathcal{F}=\sqrt{\left\langle \psi _{\mathrm{id}}\right\vert \rho
\left\vert \psi _{\mathrm{id}}\right\rangle},$
where $\left\vert \psi _{\mathrm{id}}\right\rangle $ is the output state $|g\rangle_1|0\rangle_a|0\rangle_b(\alpha|g\rangle_2+\beta|e\rangle_2+\gamma|f\rangle)_2$ of an ideal system (i.e., without dissipation, dephasing, and
crosstalk); while $\rho$ is the final density operator of the system when the operation is performed in
a realistic situation.

\begin{figure}[tbp]
\begin{center}
\includegraphics[bb=-13 585 597 818, width=14 cm, clip]{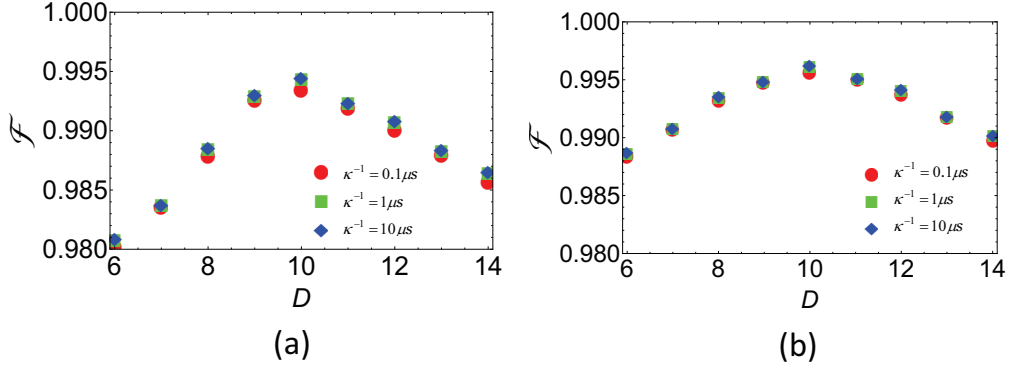}
\vspace*{-0.08in}
\end{center}
\caption{Fidelity of the state-transfer operation versus the ratio $D=\delta/g$, plotted for
$\kappa^{-1}=0.1~\mu s,1~\mu s,10~\mu s.$ (a) Case for $\alpha=\beta=\gamma=1/\sqrt{3}$.
(b) Case for $\alpha=1/\sqrt{2},$ $\beta=1/\sqrt{3}$, and $\gamma=1/\sqrt{6}$.
Here and in Figs.~4, 5, we choose $\kappa^{-1}=\kappa_a^{-1}=\kappa_b^{-1}$, $\kappa
_{a}$ ($\kappa_{b}$) is the photon decay rate of resonator $a$ ($b$). }
\label{fig:3}
\end{figure}

We now numerically calculate the fidelity of operation. Without loss of generality, consider identical
flux qutrits. For SC flux qutrits, the transition frequency between two neighbor levels is 1 to 30 GHz. Thus, we choose $\omega_{eg}/2\pi=3.5$ GHz and $\omega _{fg}/2\pi=8.8$ GHz. In addition, we set $\delta/2\pi=1.0$ GHz and $\Delta/2\pi=0.8$ GHz. For the setting here, we have $\omega_{a}/2\pi=2.5$ GHz, $\omega _{b}/2\pi=8.0$ GHz and $\Delta_{ab}=5.5$ GHz. Other parameters used in the numerical simulation are: (i) $\Omega /\left( 2\pi \right) =100$ MHz
(available in experiments \cite{s19,s36}); (ii) $\gamma^{-1} _{j,\varphi e}=\gamma ^{-1}_{j,\varphi f}=2~\mu s;$ (iii) $\gamma^{-1} _{eg}=\gamma ^{-1}_{fe}=\gamma ^{-1}_{fg}=5~\mu s$. Here we consider a rather conservative case for the decoherence times of flux qutrits~\cite{Pop,s13}.

In Fig.~3(a,b), we will consider the case of $\alpha=\beta=\gamma=1/\sqrt{3}$ and $\alpha=1/\sqrt{2},$ $\beta=1/\sqrt{3}$, $\gamma=1/\sqrt{6}$.  For the parameters chosen above, the fidelity versus $D=\delta/g$ is plotted in Fig.~3(a,b) for $\kappa^{-1}=0.1~\mu s,1~\mu s,10~\mu s.$ Here, we choose $\kappa^{-1}=\kappa_a^{-1}=\kappa_b^{-1}$. Fig.~3(a) [3(b)] shows that when $D=10$, the fidelity value is the optimum and a high fidelity $99.34\%$, $99.44\%$, $99.45\%$ ($99.56\%,~99.61\%,~99.62\%$) can be obtained for the resonator coherence times $0.1~\mu s,1~\mu s,10~\mu s,$ respectively. Thus, we choose $D=10$ in the following analysis.
For $D=10$, we have $g/2 \pi= 100$ MHz and $\mu/2 \pi= 80 $ MHz because of Eq.~(8). In this case, the estimated operation
time is $\sim0.19~\mu s$. The values of $g$ and $\mu$ here are readily available in experiments \cite{s37}. In Figs.~3-5, we choose $g_{ab}=0.1g$, which is readily satisfied in experiments \cite{s38}.

For the resonator frequencies and the values of $\kappa^{-1}\in\{0.1~\mu s,1~\mu s,10~\mu s\}$ used in the numerical calculation, the required quality factors \{$Q_a$,$Q_b$\} for the two resonators are $\{1.57\times10^{3},5.02\times10^{3}\}$,$\{1.57\times10^{4},5.02\times10^{4}\}$,$\{1.57\times10^{5},5.02\times10^{5}\}$, respectively. Note that SC coplanar waveguide
resonators with a loaded quality factor $Q\sim 10^{6}$ \cite{s40,s41} or with
internal quality factors above one million ($Q>10^{7}$)
have been previously reported \cite{s42}. Recently, a SC microwave resonator
with a loaded quality factor $Q\sim 3.5\times10^{7}$ has been
demonstrated in experiments \cite{s43}.
Fig.~3(a,b) shows that the state-transfer operation can be high-fidelity performed assisted by the low-$Q$ resonators.

\begin{figure}[tbp]
\begin{center}
\includegraphics[bb=-7 328 580 748, width=11 cm, clip]{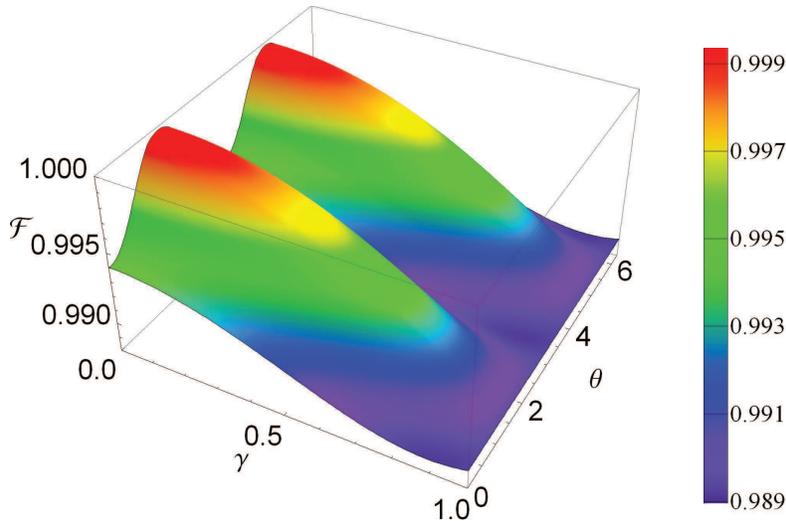}
\vspace*{-0.08in}
\end{center}
\caption{Fidelity versus the $\gamma$ and $\theta$, plotted for $D=10$ and
$\kappa^{-1}=0.1~\mu s$.}
\label{fig:4}
\end{figure}

\begin{figure}[tbp]
\begin{center}
\includegraphics[bb=-15 409 554 781, width=11 cm, clip]{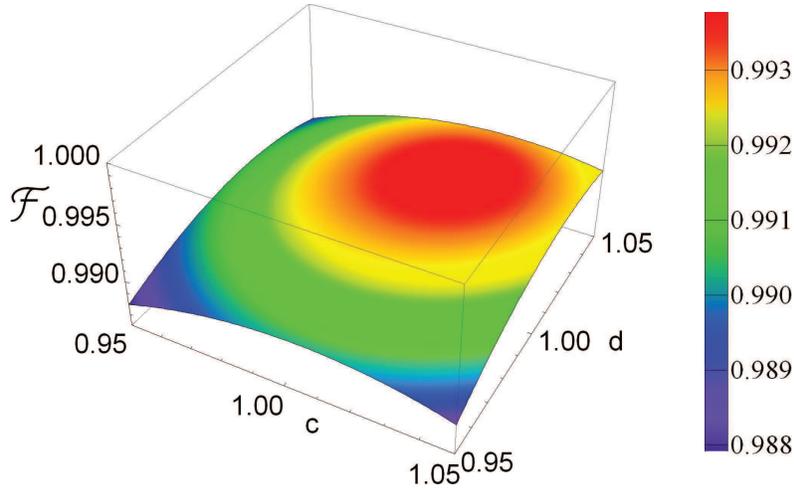}
\vspace*{-0.08in}
\end{center}
\caption{Fidelity versus the $c$ and $d$, plotted for $D=10$ and
$\kappa^{-1}=0.1~\mu s$. Here, $c=g_2/g_1$ and $d=\mu_2/\mu_1$, with $g_1=g$ and $\mu_1=\mu$.}
\label{fig:5}
\end{figure}

In Fig.~4, we numerically calculate the fidelity of state-transfer operation for a large number of randomly chosen
initial states and give the resulting distribution of output
fidelities. We set $\alpha=\sqrt{1-\gamma^2}\sin\theta$ and $\beta=\sqrt{1-\gamma^2}\cos\theta$
with $\gamma \in[0,1]$ and $\theta \in[0,2\pi]$. Figure 4
shows the fidelity versus $\gamma$ and $\theta$, which is plotted for $D=10$ and
$\kappa^{-1}=0.1~\mu s$. From Fig.~4, one can see that for $\gamma \in[0,1]$ and $\theta \in[0,2\pi]$, the fidelity can be greater than $98.9\%$.
Moreover, we calculate the average fidelity $\sim$ $99.2\%$
for $\gamma \in[0,1]$ and $\theta \in[0,2\pi]$. Figure~4 also displays that the state-transfer operation can be high-fidelity performed assisted by the low-$Q$ resonators.

Without losing the generality, we consider the case of inhomogeneous qutrit-resonator with $g_1\neq g_2$ and $\mu_1\neq \mu_2$.
For simplicity, we consider the case for $\alpha=\beta=\gamma=1/\sqrt{3}$ in Fig.~5. We set $g_1=g, \mu_1=\mu$ but $g_2=cg ,\mu_2=d\mu$, with $c,d \in[0.95,1.05]$.
Figure~5 shows the fidelity versus $c$ and $d$, which is plotted by
choosing $D=10$ and
$\kappa^{-1}=0.1~\mu s$. From Fig.~5, one
can see that for $c \in[0.95,1.05]$ and $d \in[0.95,1.05]$, the fidelity can be greater than $98.8\%$. Our numerical simulation indicates that the
high-fidelity implementation of a qutrit-to-qutrit state transfer is feasible
with current circuit-QED technology.

\section{Conclusion}
We presented a scheme to transfer an unknown qutrit state in circuit QED.
Our present proposal differs from the Refs.~\cite{14Bayat,s44}.
First, compared with Ref.~\cite{14Bayat}, no measurement is required in our proposal.
Second, compared with the previous method in Ref.~\cite{s44}, our proposal can be realized with the high fidelity assisted by the low-$Q$ resonators due to the virtually excited resonator photons during the operation time.

As shown above, because the resonator photons
are virtually excited for during the operation time, the decoherences caused by the resonator decay and
the unwanted inter-resonator crosstalk are greatly suppressed. In addition, our approach is quite
general because it can be applied to accomplish the same task with other solid-state qutrits coupled to circuit resonators. Our numerical simulation shows that the high-fidelity implementation of the qutrit-to-qutrit quantum state transfer is feasible with the state-of-the-art circuit QED technology. These will contribute to raising experimental enthusiasm for transferring an unknown qutrit state in the near future.

\section*{Acknowledgement}

This work was supported by the National Natural Science
Foundation of China, under Grant No.11375036, the Xinghai Scholar
Cultivation Plan and the Fundamental Research Funds for the Central
Universities under Grant No. DUT15LK35.

\end{document}